\def\gapprox{\lower .7ex\hbox{$\;\stackrel{\textstyle >}{\sim}\;$}}
\def\lapprox{\lower .7ex\hbox{$\;\stackrel{\textstyle <}{\sim}\;$}}
\def\bfr{{\bf r}}

\def\qbar{\overline{q}}
\newcommand{\be}{\begin{equation}}
\newcommand{\ee}{\end{equation}}
\newcommand{\bea}{\begin{eqnarray}}
\newcommand{\eea}{\end{eqnarray}}

\documentstyle[12pt,epsfig,a4]{article}
\newlength{\dinwidth}
\newlength{\dinmargin}
\begin{document}
\titlepage
\begin{flushright}
DTP/98/50\\
July 1998
\end{flushright}

\vspace*{1in}
\begin{center}
{\Large \bf Saturation Effects in Deep Inelastic Scattering \\
at low $Q^2$ and its Implications on Diffraction}\\
\vspace*{0.5in}
K. \ Golec-Biernat\footnote{On leave from H.\ Niewodniczanski 
Institute of Nuclear Physics, 
Department of Theoretical Physics, ul. Radzikowskiego 152, Krakow, Poland.} 
and M. W\"usthoff \\
\vspace*{0.5cm}
{\it Department of Physics, University of Durham, Durham DH1 3LE, UK} \\
\end{center}
\vspace*{2cm}
\centerline{(\today)}

\vskip1cm 
\begin{abstract}
We present a model based on the concept of saturation for small $Q^2$ and 
small $x$. With only three parameters we achieve a good description 
of all Deep Inelastic Scattering data below $x=0.01$. This includes
a consistent treatment of charm and a successful extrapolation into
the photoproduction regime. The same model leads to 
a roughly constant ratio of diffractive and inclusive cross section. 
\end{abstract}

\newpage
\section{Introduction}
\label{sec:1}
The basic concept of saturation in Deep Inelastic Scattering (DIS) 
is related to the transition from high to low $Q^2$ as one observes
in the total $\gamma^* p$-cross section. This type of saturation occurs
when the photon wavelength $1/Q$ reaches the size of the proton. 
We will include another aspect of saturation in our paper
which is inherent to DIS at small $x$ (small-$x$ saturation). In this regime
the partons in the proton form a dense system with mutual interaction
and recombination which also leads 
to the saturation of the total cross section \cite{GLR}.
Both aspects of saturation are closely linked to confinement
and unitarity. While the latter might be approached perturbatively
\cite{Bar}, the first is genuinely nonperturbative. The approach
we choose here can be called QCD-inspired phenomenology and follows
the line of refs.\cite{Mue,NIK1,NIK3}. It is in spirit 
most similar to the ideas of the analysis \cite{Genya}.

The basis for our approach is the fact
that the photon splits up into a quark-antiquark pair (dipole),
far upstream the proton target, which then scatters on the proton. 
In the pure perturbative
regime the reaction is mediated by single-gluon exchange which changes
into multi-gluon exchange when the saturation region is approached.
The latter process 
can be interpreted as the interaction with a 'semiclassical field' 
\cite{BuHe}. Most important for us is the fact that
the mechanism leading to the dissociation of the photon and the 
subsequent scattering
can be factorized and written in terms of a photon wave function
convoluted with a quark-antiquark cross section $\hat{\sigma}$ \cite{JEFF}. 
In our analysis this cross section has the simple form 
$\hat{\sigma}=\sigma_0\{1-exp[-r^2/(4R_0^2(x))]\}$
where $r$ denotes the separation between the quark and antiquark and
$R_0^2$ is the $x$-dependent saturation scale: $R_0^2(x)=(x/x_0)^\lambda$. 
The functional
form of $\hat{\sigma}$ can be different, important is, however, 
that for small $r$ it is proportional to $r^2$ (colour transparency)
while for large $r$ the cross section approaches a constant value.
The latter behaviour ensures saturation.
The crucial element in our analysis is the assumption that
the saturation scale $R_0$ depends on $x$ in such a way that  
with decreasing
$x$-Bjorken one has to go to smaller distances (higher $Q^2$)
in order to resolve the dense parton structure of the proton.
The boundary in the $(x,Q^2)$-plane along which saturation sets in
is described by the ``critical line'', $Q^2=1/R_0^2(x)$.

The recent low-$Q^2$ data from HERA \cite{H1,ZEUS}
have triggered significant theoretical
activity. In one class of models \cite{BK,MarRysSta} the description
of the data is achieved by combining the nonperturbative 
vector meson contribution 
(VDM-model) with contributions based on perturbative QCD.
Both models do not use the concept of small-$x$ saturation.

Another class of models \cite{DL,Rue,CKMS} is based on Regge theory. 
The total cross section $\sigma^{\gamma^* p}$ is the sum of 
contributions from different Reggeons. In particular in ref.\cite{DL}
the leading behaviour is given by the sum of two Pomeron contributions
with ``hard'' and ``soft'' intercepts. Again the concept of the small-$x$
saturation is not implemented in these analyses. The claim in
ref.\cite{DL} is that saturation at present energies is not needed. 
However,
in the true high energy asymptotics something has to happen. The hard Pomeron
as proposed in ref.\cite{DL} would eventually 
overtake and strongly violate unitarity.
Other approaches like those in refs.\cite{GVDM,BuHa} have imposed 
a logarithmic behaviour in $x$ from the very beginning 
and thus do not violate unitarity.

The strategy we adopt in our analysis is the following. 
We determine the three free parameters of our model mentioned earlier,
 $\sigma_0,\lambda$ and $x_0$, by fits to all existing DIS data for 
$x \le 0.01$.
We then study the obtained parameterization in the photoproduction region, 
where a non-zero quark mass is required to achieve a 
finite cross section. The quark mass
is chosen such that the photoproduction cross section is in rough
agreement with the data without having them included in our fit.
We found that the effective slope 
$\lambda_{eff}$ of the cross section ($\sigma\sim (W^2)^{\lambda_{eff}}$)
interpolates between the ``soft'' Pomeron value $\approx 0.08$ and the ``hard''
value $\approx 0.29$. We would like to point out 
that the ``soft'' value is simply 
a result of the extrapolation of our fits to the  photoproduction region.
After a first fit with only light quark flavours we perform a 
second fit which includes charm. We found again a good description of all
inclusive DIS data and in addition the correct relative charm contribution.
The model we use can be straightforwardly applied to DIS diffractive
processes leading to the interesting result that the ratio of the
diffractive and inclusive cross section is roughly constant as
a function of $x$ and $Q^2$. 

The content of the paper is the following.
In section 2 we introduce the theoretical details 
of the model and discuss its qualitative features. We then derive
a simplified parameterization which illustrates
the concept of the critical line. This parameterization
can also serve for a quick fit to the 
data. In Section 3 we describe our two fits and discuss their
physical implications. Section 4 is devoted to the study of diffraction
based on our model and section 5 contains conclusions.

\section{Models for Saturation}
\label{sec2}
 
For small values of the Bjorken variable $x$
the photon wave function formalism has been established
as useful tool for calculating deep inelastic and related
diffractive cross sections for 
$\gamma^* p$ scattering \cite{Mue,NIK1,JEFF}. It allows
to separate between the wave function of the photon which
describes the dissociation of the photon into a 
quark-antiquark pair and the interaction of the quark-antiquark
pair with the target. The photon wave function constitutes the calculable
part of the process whereas the remainder is substantially influenced
by nonperturbative contributions and needs to be modelled.
The corresponding diagram is shown in Fig.~1. We work 
in a frame where the  photon with momentum $q$ and the proton with momentum
$p$ are collinear. Accordingly, the distribution of  
the quark-antiquark pair is given in terms of $z$ and $(1-z)$, 
the momentum fraction with respect
to $q$, and the relative transverse separation ${\bf r}$. 
For transverse $({T})$ and longitudinally $({L})$ polarized photons
the $\gamma^* p$-cross section takes the form \cite{NIK1,JEFF}
\begin{figure}[t]
   \vspace*{-1cm}
    \centerline{
     \epsfig{figure=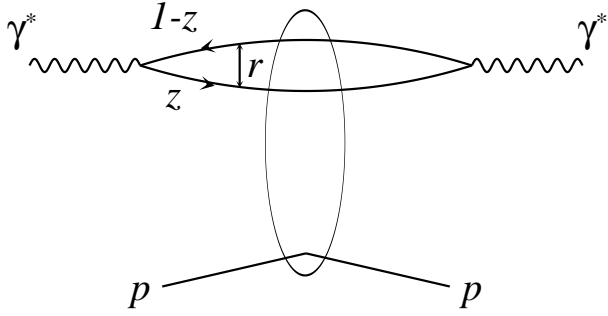,width=10cm}
               }
    \vspace*{-0.5cm}
\caption{Diagrammatical representation of the basic process as discussed
in the text.}
\label{fig1}
\end{figure} 
\be
\label{eq:1}
\sigma_{T,L}(x,Q^2)\:=\:\int \!d\,^2{\bf r}\! \int_0^1 \!dz \:  
\vert \Psi_{T,L}\,(z,\bfr) \vert ^2 \: \hat{\sigma}\,(x,r^2)\;,
\\
\ee
where $W^2=(p+q)^2$, $Q^2=-q^2$ and $x=Q^2/(W^2+Q^2)$. The squared photon wave 
function $\Psi_{T,L}$ is given by
\be
\label{eq:2}
\vert \Psi_{T}\,(z,\bfr) \vert^2  \:=\:
\frac{6\, \alpha_{em}}{4\,\pi^2 } \: \sum_{f}\,e_f^2 \:
\Big\{\,[\,z^2+(1-z)^2\,]\:\epsilon^2\,K_1^2\,(\epsilon\, r) 
\,+\, m_f^2\: K_0^2\,(\epsilon\, r)\,\Big\}
\ee
and
\be
\label{eq:3}
\vert \Psi_{L}\,(z,\bfr) \vert^2  \:=\:
\frac{6\, \alpha_{em}}{4\,\pi^2 } \:\sum_{f}\,e_f^2 \:
\Big\{\,4\,Q^2\,z^2\,(1-z)^2\,K_0^2\,(\epsilon\, r)\,\Big\}\;,
\ee
for the transverse and longitudinal photons, respectively. In the above
formulae
\be
\label{eq:4}
\epsilon^2 \:=\:  z\,(1-z)\,Q^2\,+\,m_f^2\;.
\ee
$K_0$ and $K_1$ are Mc Donald functions
and the summation is performed over the quark flavours.  
The $\gamma^* p$ cross sections are related to 
the structure function $F_2$ in the following way
\be
\label{eq:5}
F_2\,(x,Q^2)\,=\,F_{T}\,(x,Q^2)\,+\,F_{L}\,(x,Q^2) 
\ee
and
\be
\label{eq:6}
F_{T,L}\,(x,Q^2)\,=\,
\frac{Q^2}{4\,\pi^2\,\alpha_{em}}\;\sigma_{T,L}(x,Q^2)\;.
\ee

The interaction of the  $q \qbar$ pair  with 
the proton is described by the dipole cross section $\hat{\sigma}\,(x,r^2)$ 
which is modelled in our analysis. The most crucial element 
is the adoption of the $x-$dependent radius 
\be\label{eq:7}
R_0\,(x)\,=\, \frac{1}{Q_0}\;\Bigg( \frac{x}{x_0} \Bigg)^{\lambda/2}\;,
\ee
which scales the quark-antiquark separation $r$ in the dipole cross section
\be\label{eq:8}
\hat{\sigma}\,(x,r^2)\;=\; \sigma_0\; g\,(\hat{r}^2)\;,
\ee 
with 
\be\label{eq:8a}
\hat{r}\;=\; \frac{r}{2\,R_0(x)}\;.
\ee
$Q_0=1~GeV$ in (\ref{eq:7}) sets the dimension.
The function $g$ in (\ref{eq:8}) is not completely constrained. Important,
however, is the quadratic rise  at small $\hat{r}$ and the flattening
off at large $\hat{r}$. The latter behaviour provides {\it saturation} of the
cross section (\ref{eq:1}), i.e. $\sigma^{\gamma^* p}=\sigma_T+\sigma_L\sim
const$ for small $Q^2$. At small $\hat{r}$, on the other hand, 
we have a simple scaling behaviour (colour transparency), $\sigma^{\gamma^* p}
\sim 1/Q^2$, combined with a power-like dependence of (\ref{eq:8}) on $x$
as typically observed in deep inelastic scattering.  
We choose the following simple Ansatz for the function $g$
\be\label{eq:9}
g\,(\hat{r}^2)\;=\;1\,-\,e^{-\hat{r}^2}\;.
\ee
This Ansatz reminds of eikonalization. It should be mentioned, however, 
that a complete eikonal treatment requires 
the incorporation of a target profile function. The form (\ref{eq:9})
would mean in this context that the gluonic density in the proton
is evenly distributed
over a certain area within a sharp boundary and zero beyond.
A more sophisticated treatment including a realistic profile function
can be found in ref.\cite{NIK3,Genya}. The corresponding 
result in these references can be roughly approximated by
\be\label{eq:10}
g\,(\hat{r}^2)\;=\;\ln(1+\hat{r}^2)
\ee 
and shows a logarithmic growth at large distances. 

For small $z$ one can as well think of a logarithmic modification 
as is motivated by the single gluon exchange (see ref.\cite{NIK1}) 
\be\label{eq:11}
g\,(\hat{r}^2)\;=\;\hat{r}^2\ln\left(1+\frac{1}{\hat{r}^2}\right)\;\;.
\ee

Both alternatives (\ref{eq:10}) and (\ref{eq:11}) can be 
easily implemented in our formalism which we present in the forthcoming
sections. 
It turns out, however, that they are disfavoured in our analysis
 and we will not  discuss  them in detail. This does not mean 
that we contradict their physical implications. It might be that a different
approach could  be  reconciled with the models
(\ref{eq:10}) or (\ref{eq:11}).

In our analysis we fit the three parameters, $\sigma_0$, $x_0$ and $\lambda$, 
of the dipole cross section (\ref{eq:8}) and (\ref{eq:9}).
Before going into the details of the fits  
it is illuminating to perform a qualitative analysis of the behaviour of
the cross section (\ref{eq:1}) in our model.

\subsection{Qualitative Analysis} \label{sec3}

In the following discussion we focus on the cross section $\sigma_T$ 
in (\ref{eq:1}) which dominates over $\sigma_L$. 
We also neglect for simplicity the quark mass.  
The important point in the qualitative analysis is the behaviour 
of $K_1(\epsilon\, r)$ in (\ref{eq:2}) for small values of $\epsilon\, r$:
\be
\label{eq:12}
K_1(\epsilon\, r)\,\sim\,\frac{1}{\epsilon\,r}\; .
\ee
For large values of $\epsilon\,r$ the function
$K_1$ is exponentially suppressed.
Thus in order to obtain the dominant contribution 
we perform the integration in (\ref{eq:1}) for
$\epsilon\,r  < 1$.

The photon virtuality introduces the scale 
$1/Q$ for the transverse dimension
of the $q\qbar$ pair. A pair is considered  ``small'' when  the condition 
$r < 1/Q$ is fulfilled and ``large'' when $r > 1/Q$.
Let us analyse the contribution to (\ref{eq:1}) coming from small 
pairs for which the condition $\epsilon\, r =\sqrt{z (1-z)}\,r\, Q < 1$ 
is satisfied for all values of $z$.
In the case $1/Q \ll R_0$, shown in Fig.~2a, the size of the small $q\qbar$
pairs is much smaller than the saturation radius and  
$\hat{\sigma}(r)\;\sim\; \sigma_0\, \hat{r}^2/R_0^2$.
The cross section (\ref{eq:1})  exhibits the following behaviour
\be
\label{eq:13}
\sigma_T\: \sim \:
\frac{\sigma_0}{R_0^2}\,  
\int dz
\int\limits_0^{1/Q^2} d\,r^2\,\epsilon^2\,
\Bigg(\frac{1}{\epsilon^2\,r^2}\Bigg)\,\hat{r}^2\;
\; \sim \; \frac{1}{Q\,^2}\;
\frac{\sigma_0}{R^2_0}
\ee
where the integration over $z$ could be factored after the cancellation
of the $\epsilon$ factors. Thus for constant $x$
the cross section (\ref{eq:13}) exhibits the familiar short distance
scaling behaviour, i.e the corresponding structure function
$F_2(x,Q^2)$ is roughly constant in $Q^2$.

\begin{figure}
   \vspace*{-1cm}
    \centerline{
     \epsfig{figure=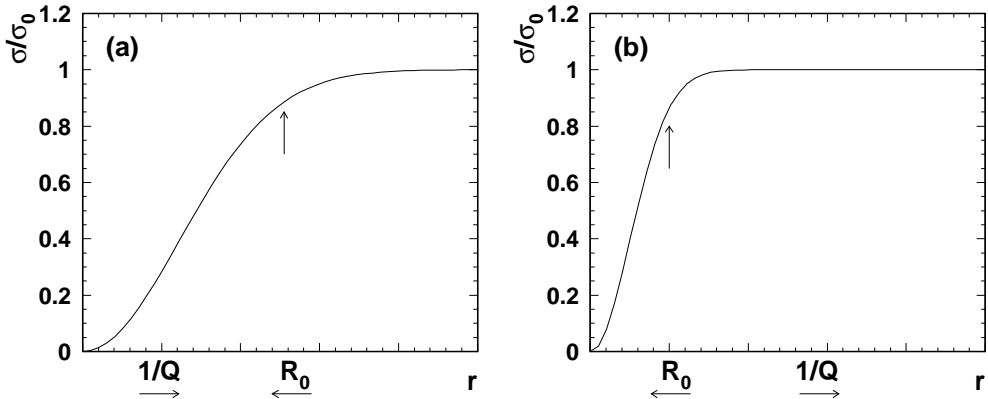,width=15cm}
               }
\label{fig2}               
    \vspace*{-0.5cm}
\caption{ The profile of the dipole cross section for different $Q$. The small
arrows below the figure show how the indicated parameters 
change when $Q$ decreases (for $W^2$  fixed).}
\end{figure}

Let us now analyse the situation shown schematically in Fig.~2b
in which the size of the $q\qbar$ pair is bigger than the saturation
radius $1/Q > R_0$. For the small pair contribution to $\sigma_T$
we now obtain
\be
\label{eq:14}
\sigma_T\: \sim \:  \int\limits_0^{R_0^2} d\,r^2\,
\Bigg(\frac{1}{r^2}\Bigg)\, \sigma_0\,\frac{\,r^2}{R^2_0}\; + \;
\int\limits_{R_0^2}^{1/Q^2} d\,r^2\, 
\Bigg(\frac{1}{r^2}\Bigg)\, \sigma_0\;
\sim \; \sigma_0 \,+\, 
\sigma_0\;\log\,\bigg(\frac{1}{Q^2\,R_0^2}\,\bigg)\;.
\ee
With respect to the power behaviour in $Q^2$ the cross section 
can be viewed as 
being constant. The potential divergence due to the logarithm will be 
regulated by
the quark mass in our full analysis. The actual value of the mass
plays an important role in the description of the photoproduction region.

A similar analysis can be performed for ``large'' pairs for which $r > 1/Q$.
The integration condition $\epsilon\,r < 1$
is now satisfied  when $z < 1/(r^2\,Q^2)$ and the $z$
integration in (\ref{eq:1}) can no longer  be factored out. It
has to be done before integrating over $r$.
The result in the end is the same as for small pairs. If the characteristic
size of the $q\qbar$ dipole is less than $R_0$ the scaling behaviour
is obtained. For $1/Q \gapprox R_0$ the cross section 
$\sigma_T$ is constant in $Q^2$. A more detailed analysis  gives
logarithmic modifications.

So far in our discussion we have assumed a constant saturation radius
which allows a smooth transition of $\sigma_T$ between the scaling region 
of large $Q^2$ and the saturation region of low $Q^2$
(low-$Q^2$ saturation). The main feature of our model, however,
is the fact that the saturation radius
$R_0$ depends on $x$ ($R_0(x)\sim x\,^{\lambda/2}$ with $\lambda>0$). 
In this way we have introduced 
another kind of saturation which can be called the small-$x$ saturation. 
In terms of the parton picture it is closely related to 
the saturation of the gluon density \cite{Genya2}. 
An important consequence is
that for fixed $W$ the radius $R_0$ becomes $Q$ dependent 
which makes the saturation more dynamical.
As illustrated in Fig.~2 both $R_0$ and $1/Q$ move towards
each other, at a certain scale $Q=1/R_0$ they meet and then pass each other. 
Hence,
the saturation occurs at higher values of $Q^2$
than in a model with a fixed $R_0$. In addition the transition from high
to low $Q^2$ is faster. The line given by the condition
\be\label{eq:14a}
R_0^2(x)=\frac{1}{Q^2}
\ee 
will be called the {\bf critical line}. In the parton picture it would
describe the boundary of the critical density \cite{Genya2}. 
The precise 
saturation pattern is determined by the fits of the three parameters in 
(\ref{eq:8}) to the existing DIS inclusive data.

\subsection{The Mellin Transformation} \label{sec4}

In this section we briefly describe the technique we use
to perform the fits. In order to evaluate the cross section (\ref{eq:1})
we employ the Mellin transformation to factorize the 
wave function from the cross section. 
This reduces the number of numerical integrations to one and 
allows for an easier discussion of the scaling behaviour:
\bea
\label{eq:tech1}
\sigma_{T,L}(x,Q^2)&=&\int \frac{d\nu}{2 \pi}
\int \!d\,^2{\bf r} \int_0^1 \!dz \;  
\vert \Psi_{T,L}\,(z,\bfr) \vert ^2 \: \int \frac{dr'^2}{r'^2}
\left(\frac{r}{r'}\right)^{1+2i\nu}\sigma\,(\tilde{x},r'^2)
\\ \nonumber
\\ \nonumber
&=&\sigma_0\;\;\int \frac{d\nu}{2 \pi} 
\;\;\left[\left(\frac{x_0}{\tilde{x}}\right)^\lambda
\frac{Q_0^2}{Q^2}\right]^{1/2+i\nu}
\mbox{H$_{T,L}$}\left(\nu,\frac{m_f^2}{Q^2}\right)\;\;\mbox{G}(\nu)\nonumber
\eea
where
\bea\label{eq:tech2}
& &\mbox{H$_T$}\left(\nu,\frac{m_f^2}{Q^2}\right) = \sum_f e_f^2 
\;\frac{6 \alpha_{em}}{2 \pi}
\;\frac{\sqrt{\pi}}{4}\; \frac{\Gamma^2(3/2+i\nu)\Gamma(1/2+i\nu)}
{\Gamma(2+i\nu)}\;\left(\frac{Q^2+4 m_f^2}{Q^2}\right)^{1/2-i\nu}
\\ \nonumber
\\ \nonumber
&\times&
\left\{ \left[\frac{(1+3 i\nu) Q^2+(3+2 i\nu) m_f^2}{Q^2+4m_f^2}
-i\nu\left(\frac{Q^2}{Q^2+4m_f^2}\right)^2\right]\right.\; 
\mbox{$_2F_1$}\left(\frac{1}{2}
+i\nu,\frac{1}{2};\frac{3}{2};\frac{Q^2}{Q^2+4m_f^2}\right)
\\ \nonumber
\\ \nonumber
& &+
\left[(1-i\nu)\frac{Q^2}{Q^2+4m_f^2}
-\frac{3+2 i\nu}{4}\right]
\left.\;
\mbox{$_2F_1$}\left(-\frac{1}{2}+i\nu,\frac{1}{2};
\frac{3}{2};\frac{Q^2}{Q^2+4m_f^2}\right)\right\}\;
\nonumber
\eea
and
\bea\label{eq:tech3}
\mbox{H$_L$}\left(\nu,\frac{m_f^2}{Q^2}\right)&=&\sum_f e_f^2 
\;\frac{6 \alpha_{em}}{2 \pi}
\;\frac{\sqrt{\pi}}{4}\; \frac{\Gamma^3(3/2+i\nu)}
{\Gamma(2+i\nu)}\;\left(\frac{Q^2+4 m_f^2}{Q^2}\right)^{-3/2-i\nu}\\
& & \times\frac{16}{30}\;\;
\mbox{$_2F_1$}\left(\frac{3}{2}+i\nu,\frac{1}{2};\frac{7}{2};
\frac{Q^2}{Q^2+4m_f^2}\right)\nonumber\;.
\eea
In addition in (\ref{eq:tech1})
\be\label{eq:tech4}
\\ \nonumber
\mbox{G}(\nu)\;=\;\int_0^\infty \;d\hat{r}^2\; 
\left(\hat{r}^2\right)^{-3/2-i\nu}\;\,g(\hat{r}^2)\;\;\;.
\ee

In order to have the right threshold behaviour and 
a smooth transition in the limit $Q^2 \rightarrow 0$
we also modify the Bjorken variable 
\be\label{eq:tech5}
\tilde{x}\;=\;x\;\frac{Q^2+4m_f^2}{Q^2}\;\;.
\ee
The Mellin transformation for the functional form (\ref{eq:9}) of 
the function $g$ which defines the dipole cross section
in our analysis yields:
\bea\label{eq:tech6}
\mbox{G}(\nu)\;=\;-\Gamma(-\frac{1}{2}-i \nu)\;.
\eea
For the sake of completeness we also present the corresponding
expressions for the alternative models in (\ref{eq:10}) and (\ref{eq:11}),
\bea\label{eq:tech7}
{\mbox{G}}(\nu)&=&
\frac{1}{1/2+i\nu}\;\frac{\pi}{\cosh(\pi \nu)}\;
\eea
and
\bea\label{eq:tech8}
{\mbox{G}}(\nu)&=&
\frac{1}{1/2-i\nu}\;\frac{\pi}{\cosh(\pi \nu)}\;.
\eea

The main purpose of introducing a mass is to have a finite 
limit $Q^2\rightarrow 0$. However in the region $Q\gapprox 1~GeV$,
where we mainly fit the data, the mass has little effect and one might
as well consider the case with zero mass. In this case the functions
(\ref{eq:tech2}) and (\ref{eq:tech3}) reduce to
\be \label{eq:tech9}
\mbox{H$_T$}(\nu,0)\;=\;\sum_f e_f^2 
\;\frac{6 \alpha_{em}}{2 \pi}\;\frac{\pi}{16}\;
\frac{9/4+\nu^2}{1+\nu^2}\;\left(\frac{\pi}{\cosh(\pi \nu)}\right)^2\;
\frac{\sinh(\pi \nu)}{\pi \nu}\;
\frac{\Gamma(3/2+i\nu)}{-\Gamma(-1/2-i\nu)}
\ee
and
\be\label{eq:tech10}
\mbox{H$_L$}(\nu,0)\;=\;\sum_f e_f^2 
\;\frac{6 \alpha_{em}}{2 \pi}\;\frac{\pi}{8}\;
\frac{1/4+\nu^2}{1+\nu^2}\;\left(\frac{\pi}{\cosh(\pi \nu)}\right)^2\;
\frac{\sinh(\pi \nu)}{\pi \nu} \;
\frac{\Gamma(3/2+i\nu)}{-\Gamma(-1/2-i\nu)}\;\;.
\ee
The results (\ref{eq:tech9}) and (\ref{eq:tech10}) are much
simpler than in the massive case and probably familiar to those readers
who have calculated the Mellin transform of the quark-box diagram in momentum
space. The only difference is the ratio of the $\Gamma$ functions at the end
of each expression which is a residue of the Fourier transform.

We use formulae (\ref{eq:tech1}) through (\ref{eq:tech6}) in our fits 
and perform the integration over $\nu$ numerically.
The main virtue of this representation is the quick convergence of the 
integrand in (\ref{eq:tech1})
which makes the numerical integration very fast. Another virtue is
the rather simple analytical structure which allows to extract
the leading scaling behaviour of $\sigma_T$ for large as well as low $Q^2$,
as will be demonstrated in the next section.

\subsection{Simplified Parametrization and Critical Line}\label{sec5}

In this section we analyse the analytical structure of
the cross section (\ref{eq:tech1}) in the complex $\nu$-plane.
We use the formulae (\ref{eq:tech9}) and (\ref{eq:tech10}) for the
massless case. The position
and characteristics of the singularities in the complex
$\nu$-plane determine the behaviour of the analysed cross section. 
In general
the $\nu$-integration runs along the real axis, and depending on the argument
$\left(\frac{x_0}{x}\right)^\lambda\frac{Q_0^2}{Q^2}$ in eq.~(\ref{eq:tech1}) 
one can close
the contour in the upper or lower part of the complex plane. For example,
in the case of large $Q^2$ and not too small $x$
(``hard'' regime) the  mentioned
argument is less than 1 and the contour has to be closed in 
the lower plane.
The first singularity encountered is a pole at $\nu=-i/2$. Depending on the
model for the dipole cross section given by 
(\ref{eq:tech6}), (\ref{eq:tech7}) or  (\ref{eq:tech8}),
it can be a  double or triple pole. 
The model (\ref{eq:tech6}) used in our analysis leads to
a double pole which generates a logarithmic behaviour 
of the cross section (\ref{eq:tech1})
\be \label{eq:simple1}
\left(\frac{x_0}{x}\right)^\lambda\frac{Q_0^2}{Q^2}\;\;\ln\left[  
\left(\frac{x}{x_0}\right)^\lambda \frac{Q^2}{Q_0^2}\right]\;\;.
\ee
One should note that the logarithm is due to the photon wave function
(the quark-box diagram) and is related
to the splitting of a gluon into a quark-antiquark pair. The factor in front
of the logarithm arises because $1/2+i\nu=1$ in eq. (\ref{eq:tech1})
and simply reflects the basic scaling behaviour of $F_2$ combined
with a certain power behaviour in $x$.
One also recognizes
that the longitudinal contribution (\ref{eq:tech8}) has only a single pole
and therefore does not produce a logarithm. Since it is not the leading
contribution we will ignore it in the following.

In the ``soft'' regime where  $\left(\frac{x_0}{x}\right)^\lambda
\frac{Q_0^2}{Q^2}$
is larger than 1 we close the contour in the upper plane and find
the leading pole at $\nu=i/2$. For the transverse cross section together with
model (\ref{eq:tech6}) it is again a double pole leading to
the following behaviour of (\ref{eq:tech1})
\be \label{eq:simple2}
\ln\left[ \left(\frac{x_0}{x}\right)^\lambda \frac{Q_0^2}{Q^2}\right]\;\;.
\ee
We can now combine the ``hard'' and the ``soft'' terms given by
eqs. (\ref{eq:simple1}) and (\ref{eq:simple2}) into one expression
\be \label{eq:simple3}
\sigma^{\gamma^* p}(x,Q^2)\;=\;\sigma'_0\;\left\{
\ln\left[ \left(\frac{x_0'}{x}\right)^{\lambda'} \frac{Q_0^2}{Q^2}+1\right]
\;+\;\left(\frac{x_0'}{x}\right)^{\lambda'}\frac{Q_0^2}{Q^2}\;\;\ln\left[  
\left(\frac{x}{x_0'}\right)^{\lambda'} \frac{Q^2}{Q_0^2}+1\right]\right\}\;\;,
\ee
where we have added $1$ in the argument of the logarithms in order
to allow for a smooth transition between the ``hard'' and the ``soft'' regimes.
We have also introduced the  parameters with a prime 
to indicate that the functional
form should be refitted to get a good description. Although eq. 
(\ref{eq:simple3}) is a rather crude approximation of the original approach
it reproduces the main features.

As was discussed in our qualitative analysis we define the saturation
as the transition from short distances to long distance with the 
characteristic scale given by the saturation radius
$R_0(x)$. Looking at eq. (\ref{eq:simple3}) we realize that the transition 
from the ``hard'' into the ``soft'' regime occurs when
$\left(\frac{x_0'}{x}\right)^{\lambda'} \frac{Q_0^2}{Q^2}=1$. 
This equality defines
basically the same  {\bf critical line} as in our
qualitative analysis of section \ref{sec3} (see eq. (\ref{eq:14a}))
\be \label{eq:simple4}
Q^2\;=\frac{1}{Q_0^2}\;\left(\frac{x_0'}{x}\right)^{\lambda'}
\;=\;\frac{1}{R_0^2(x)}\;\;.
\ee
The precise location in the $(x,Q^2)$-plane 
and the slope of the critical line is determined
from the fits  discussed in the following section.

\section{Discussion of the Fits}\label{sec6}

We will discuss separately fits with and without charm contribution. 
As we will see
charm has a quite strong influence on the fit and causes the critical line
to move towards smaller scales. For the three light flavours we assume a
common mass of $140~MeV$ which leads to a reasonable prediction
in the photoproduction region. The general dependence  of 
the total cross section $\sigma^{\gamma^*p}$ on the quark mass is such 
that the photoproduction cross section increases logarithmically
with decreasing mass and diverges in the limit of zero mass.
Thus the quark mass plays the role of a regulator for 
the photoproduction cross section.

For our fit we use all available DIS-data below the Bjorken variable 
$x=0.01$ combining  systematic and statistical 
errors in quadrature. For the numerical evaluation we have implemented
formulae (\ref{eq:tech1}) through (\ref{eq:tech6}).

\subsection{Light Flavours}\label{sec7}

\begin{figure}
   \vspace*{-1cm}
    \centerline{
     \epsfig{figure=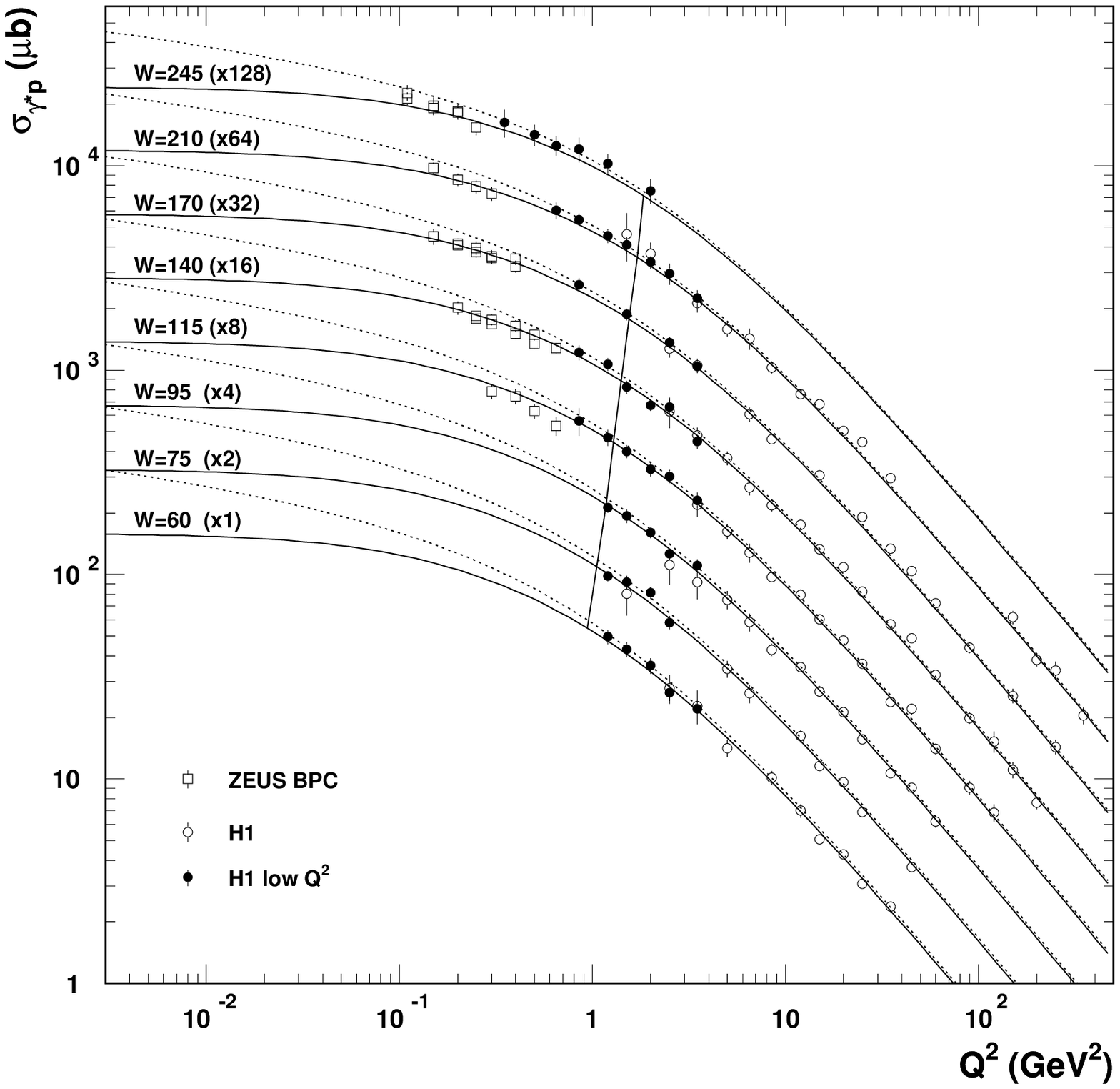,width=18cm}
               }
    \vspace*{-0.5cm}
\caption{The $\gamma^* p$-cross section for various energies. The solid lines
show the fit results with a light quark mass of 140MeV. The dotted lines show
the cross section with the same parameters but with zero quark mass.
The line across the curves indicates the position of the critical line.}
\label{fig3}
\end{figure}

The result for our first fit with only light flavours 
is listed in the first row of Table~1.
The corresponding plot in Fig.~\ref{fig3} shows only a subset of the 
data that were fitted. The most remarkable  feature
of the cross section in Fig.~\ref{fig3} is 
the turn over of the curves towards small $Q^2$ values. This illustrates
the change between the scaling and saturation region for high and
low $Q^2$, respectively, and is well reproduced by our model (solid lines). 
The critical line computed from eq.~(\ref{eq:simple4})
is plotted across all curves and indicates the onset of saturation.
For comparison we also show
the corresponding cross section when the quark mass is set to  zero 
(dotted lines). This line
demonstrates that indeed the turn over occurs basically along 
the critical line  and is not significantly influenced by the quark mass. 

\begin{figure}
   \vspace*{-1cm}
    \centerline{
     \epsfig{figure=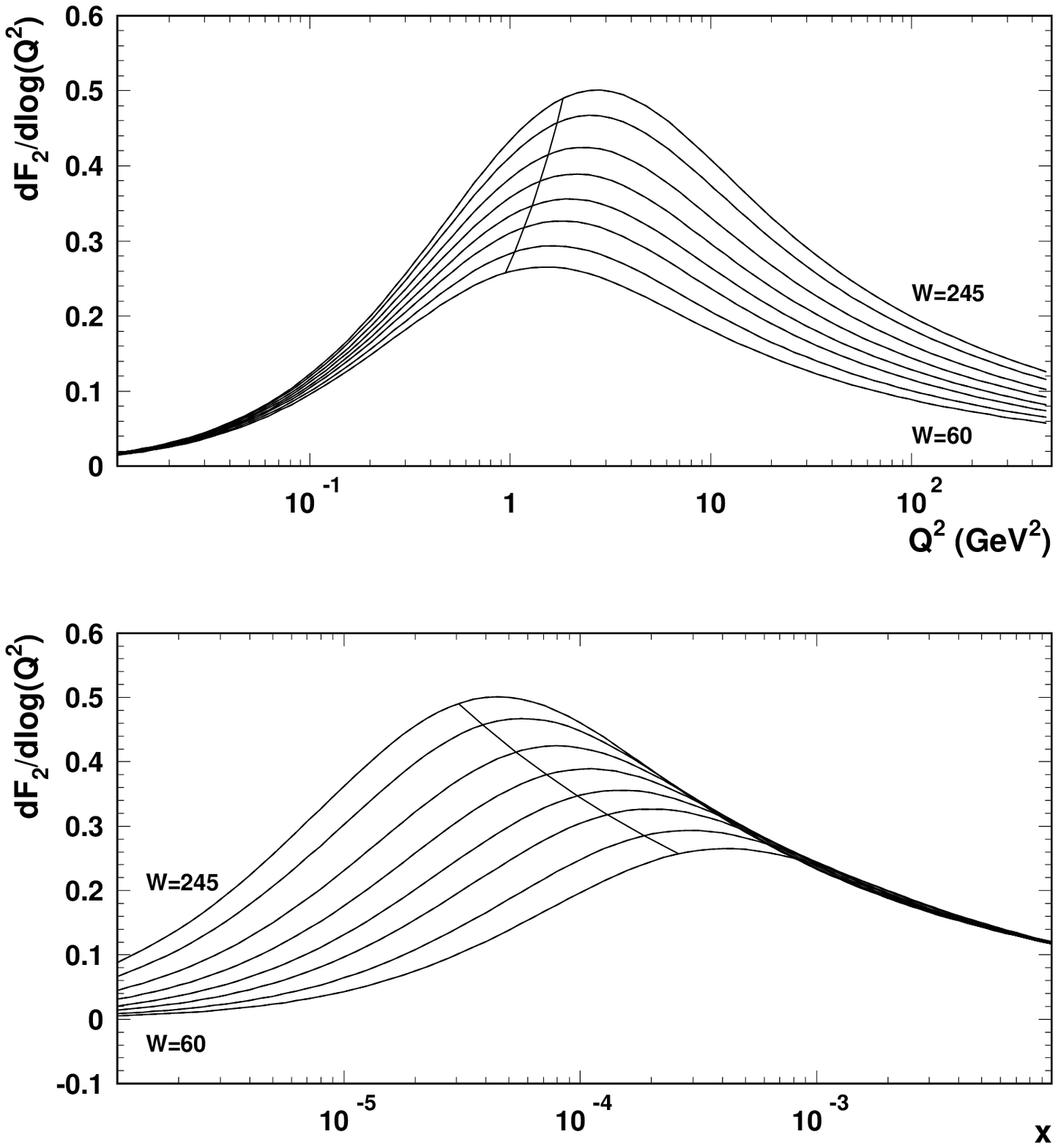,width=18cm}
               }
    \vspace*{-0.5cm}
\caption{The logarithmic $Q^2$-slope of $F_2$ for fixed energies $W$, 
plotted as a function of $Q^2$ and $x$. 
The line across the curves shows the position of the critical line.}
\label{fig4}
\end{figure}

Fig.~\ref{fig4} shows the logarithmic $Q^2$-slope of $F_2$ computed
for fixed $x$ and then plotted for fixed $W^2$ 
as a function of $Q^2$ and $x$ separately.
These plots
are inspired by an analysis which was carried out 
by ZEUS \cite{ZEUS2} to stipulate the deviation
from the conventional pQCD-approach at low values of $Q^2$ 
(see also \cite{ALLM}). 
The remarkable property of the presented plots is
a distinct maximum for each of the slopes. 
The critical line lies slightly to the left of the maxima in both plots
what might suggest that an alternative definition 
of the critical line could be introduced
as being the path along the maxima.

\begin{figure}
   \vspace*{-1cm}
    \centerline{
     \epsfig{figure=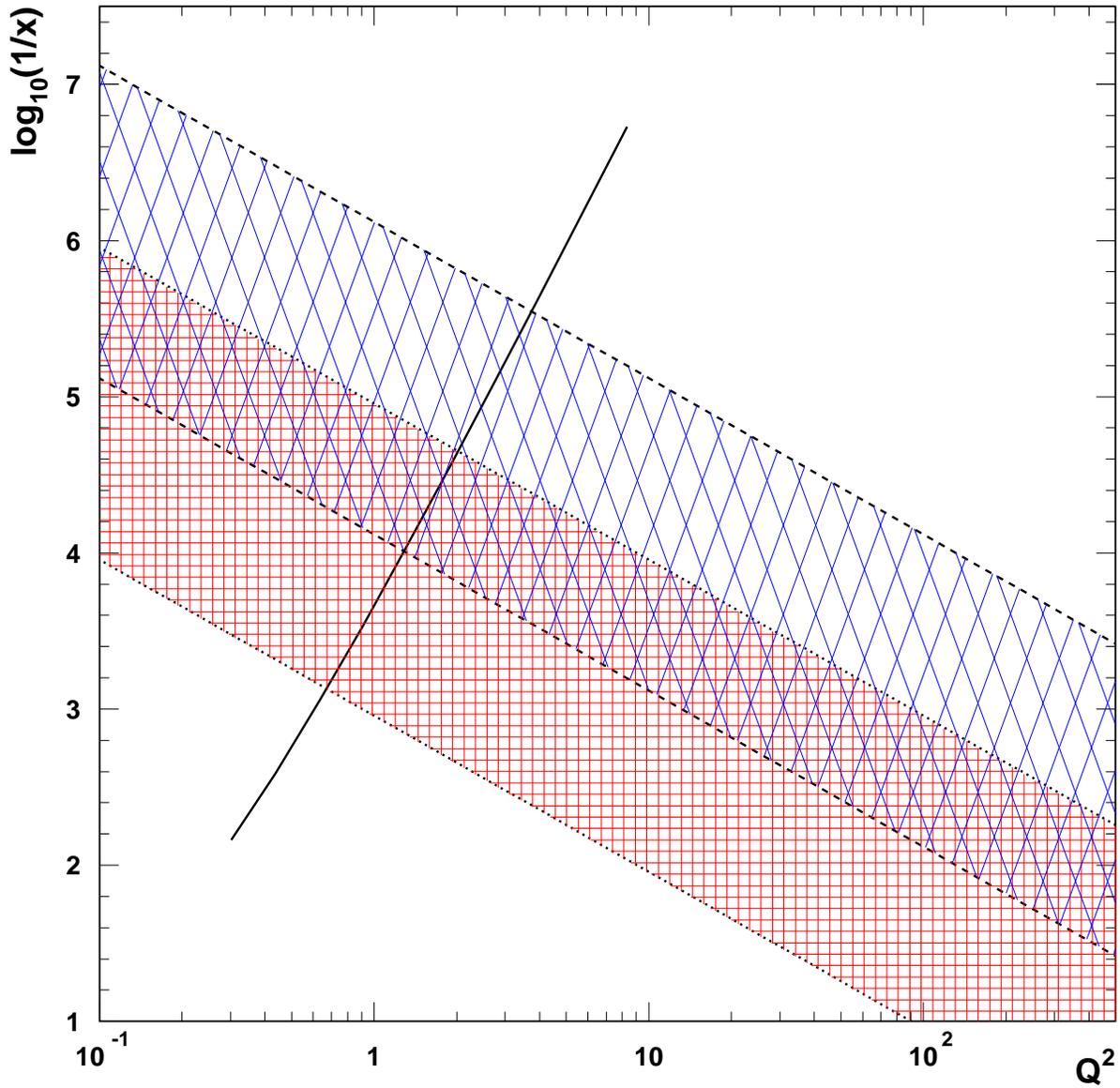,width=18cm}
               }
    \vspace*{-0.5cm}
\caption{The position of the critical line in the $(x,Q^2)$-plane.
The narrow hatched area corresponds to the acceptance region of HERA.
The wide hatched region indicates the range for a future 
$1~TeV$ $ep$-collider. The boundaries are lines
of constant $y$.}
\label{fig5}
\end{figure}

In Fig.~\ref{fig5} we show the position of the critical line obtained in our
analysis in the $(x,Q^2)$ plane. 
We observe that going along the critical line
from $x=10^{-4}$ to $x=10^{-5}$ we increase the saturation
scale from approximately $1~GeV^2$ up to $2~GeV^2$. This means that even
at a future $ep$-collider we expect only a rather small shift in the 
saturation scale. With great optimism it might go up to $3~GeV^2$ at
a $1~TeV$ machine which, nevertheless, reaches 
quite far into the perturbative regime.
\begin{table}[t]
\begin{center}
\begin{tabular}{|l|r||r|r|r|r|}
\hline
\multicolumn{5}{|c|}{\rule[-0.3cm]{0mm}{0.8cm}
\bfseries FIT RESULTS }\\
\hline
            &  $\sigma_0$ (mb) & $~~~~\lambda~~~~$&$x_0$~~~~~~~~& $~~\chi^2/(no. exp. po.)~~$ \\

\hline\hline
 no charm   & ~~23.03~~ &~~ 0.288~~ & ~~ $3.04\cdot10^{-4}$~~      &~~$441/372=1.18$~~~   \\
\hline\hline 
 with charm   & ~~29.12~~ & ~~0.277~~ &~~ $0.41\cdot10^{-4}$~~    &~~$558/372=1.50$~~~ \\
\hline
\end{tabular}
\end{center}
\end{table}

\begin{figure}
   \vspace*{-1cm}
    \centerline{
     \epsfig{figure=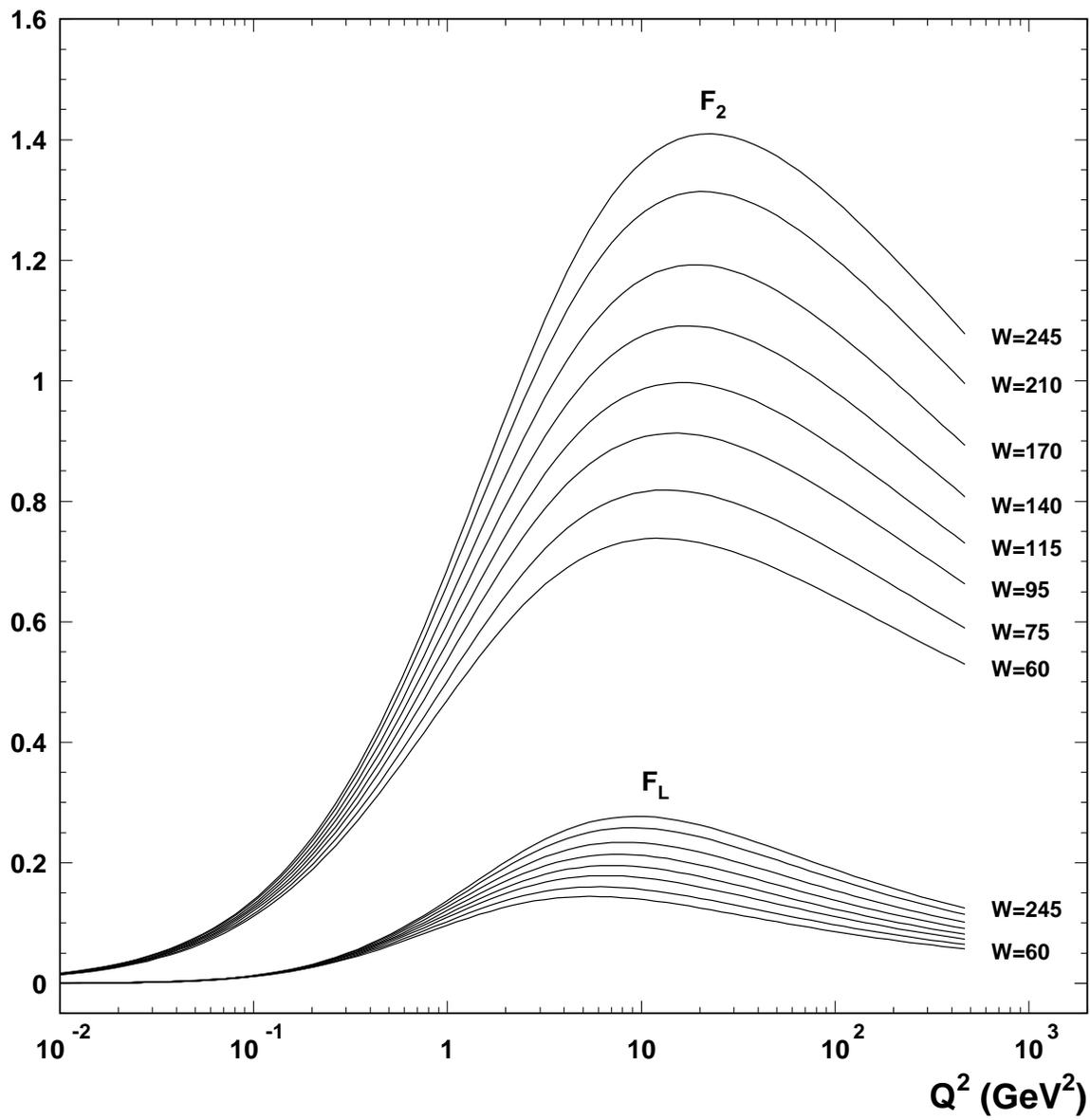,width=18cm}
               }
    \vspace*{-0.5cm}
\caption{$F_2$ and $F_L$ structure functions plotted as a function of
$Q^2$ for fixed energies $W$.}
\label{fig6}
\end{figure}

An important prediction resulting from
our  analysis is  the longitudinal structure function $F_L$ which is shown
in Fig.~\ref{fig6}. As we see  $F_L$ constitutes roughly $20\%$ 
of $F_2$ for $Q^2$ around $10~GeV^2$ which is a reasonable 
value.

\begin{figure}
   \vspace*{-1cm}
    \centerline{
     \epsfig{figure=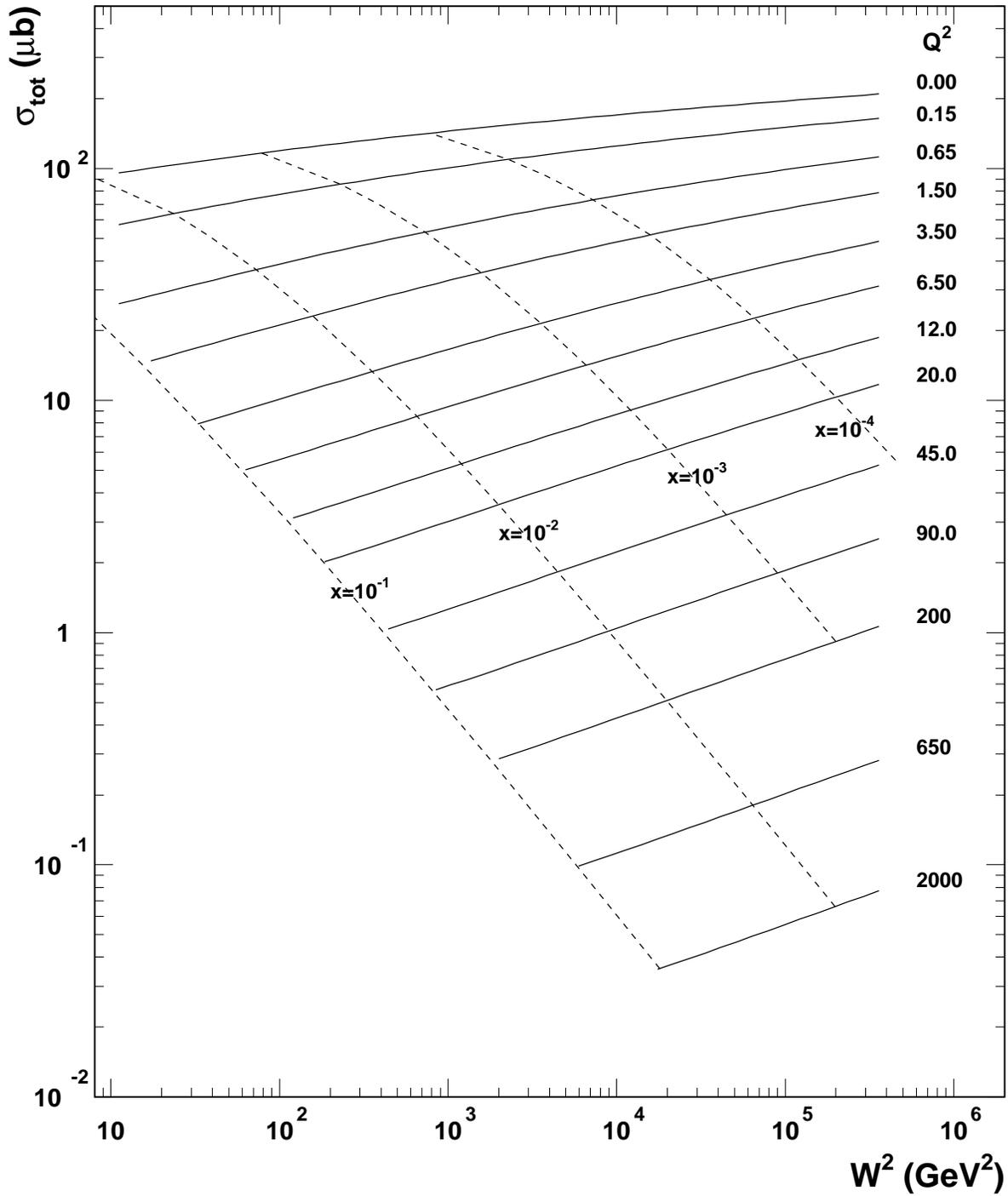,width=18cm}
               }
    \vspace*{-0.5cm}
\caption{The total $\gamma^* p$-cross section as a function of $W^2$ for
different $Q^2$-values. The dashed lines represent the lines of constant
$\tilde{x}$ as defined in eq.(\ref{eq:tech5}).}
\label{fig7}
\end{figure}

\begin{figure}
   \vspace*{-1cm}
    \centerline{
     \epsfig{figure=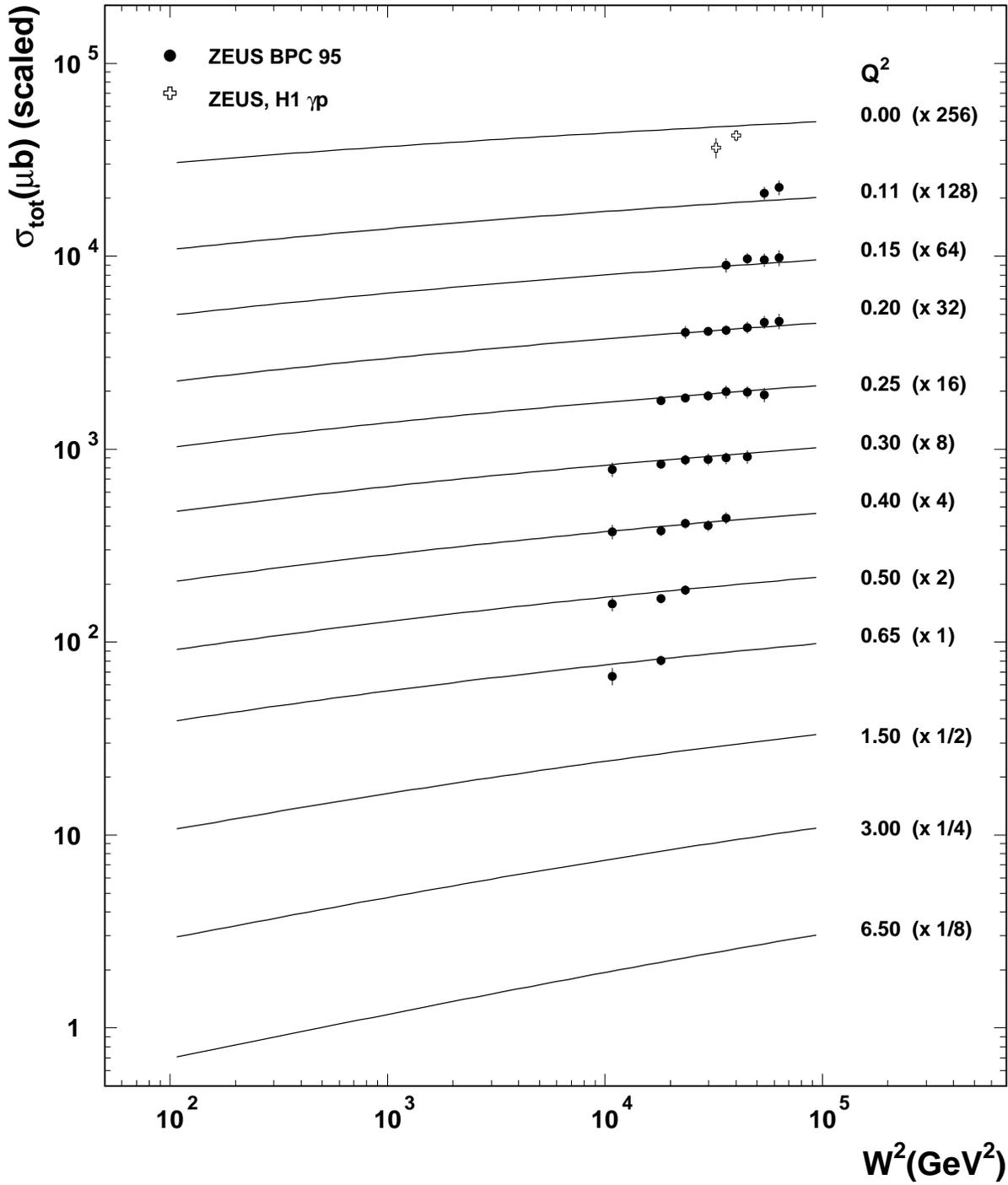,width=18cm}
               }
    \vspace*{-0.5cm}
\caption{The same cross section as in Fig.~\ref{fig7} 
with the emphasis on small-$Q^2$ values.} 
\label{fig8}
\end{figure}

We now concentrate on the $W^2$ dependence of the total cross section
$\sigma^{\gamma^*p}$ for fixed $Q^2$. We are particularly interested in 
the effective slope of  $\sigma^{\gamma^*p}$ as a function of $W^2$.
This dependence is shown in Fig.~\ref{fig7} in a broad range of
$Q^2$ values and in Fig.~\ref{fig8} with a particular interest for
smaller values of $Q^2 < 6.5~GeV^2$.
Moving from high to low $Q^2$ we see how the slope of $\sigma^{\gamma^*p}$
flattens off indicating the low-$Q^2$ saturation. One can
also recognize a slight curvature in particular for $Q^2=0$. This
is associated with saturation in energy. 

\begin{figure}
   \vspace*{-1cm}
    \centerline{
     \epsfig{figure=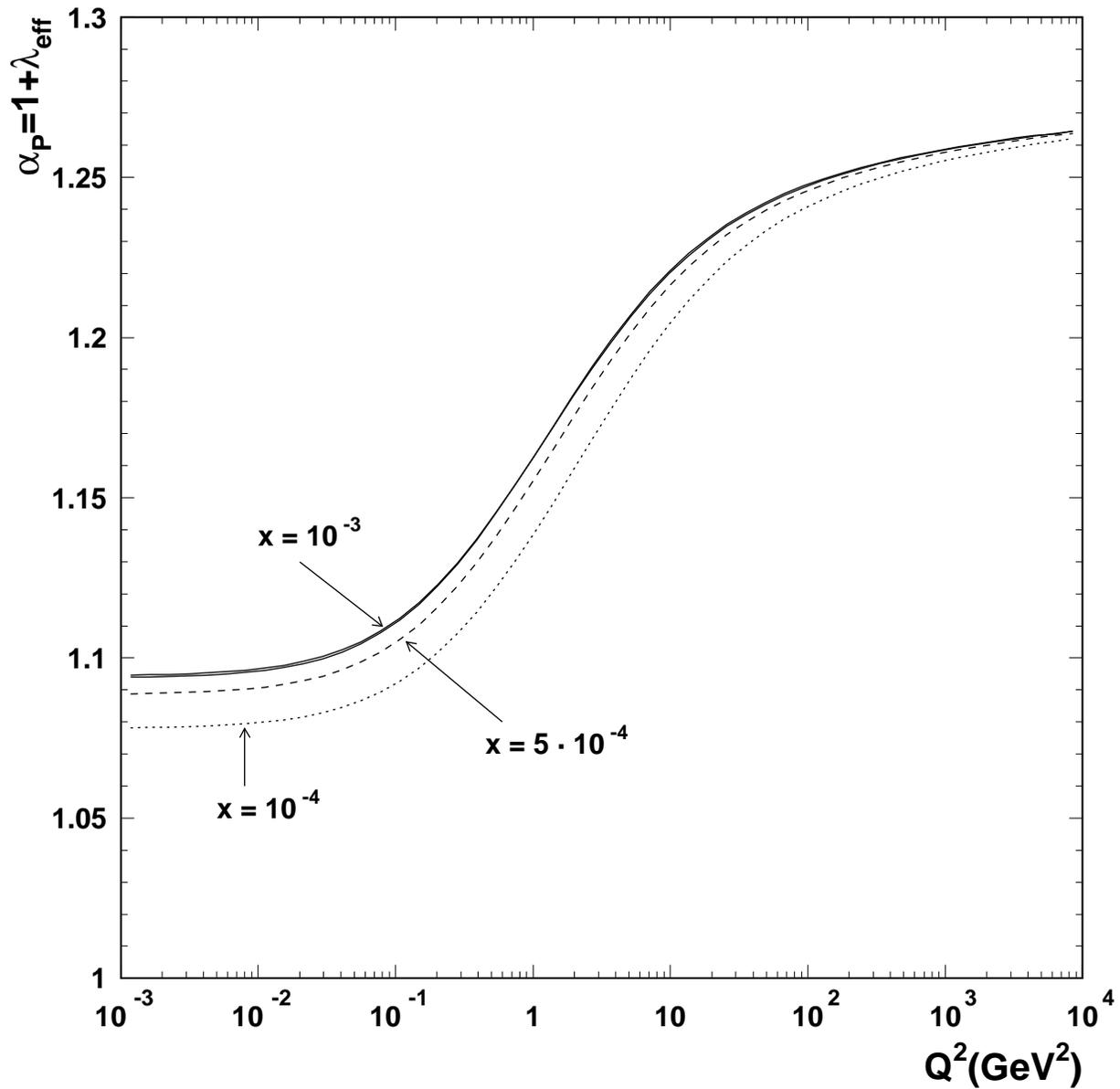,width=18cm}
               }
    \vspace*{-0.5cm}
\caption{The effective slope in $W^2$ of the total cross section, 
$\lambda_{eff}$, plotted as a function of $Q^2$ for fixed $\tilde{x}$. 
The solid line appears twofold
due to two different methods of calculation. One is based on the analytical
expression (\ref{eq:fit1}), the second is calculated  
from fig.\ref{fig7} by taking a pair of points along the lines of
constant $\tilde{x}=10^{-2}$ and $10^{-4}$.}
\label{fig9}
\end{figure}

A more explicit way of exposing
saturation is achieved by plotting the effective slope $\lambda_{eff}$
computed from the relation
\be\label{eq:fit1}
\lambda_{eff}\;=\;\frac{\partial \ln (\sigma^{\gamma^*p})}
{\partial \ln (W^2)}
\ee
as a function of $Q^2$ for fixed values of $\tilde{x}=(Q^2+4m_f^2)/(Q^2+W^2)$.
This means that we move along the dashed lines shown in Fig.~\ref{fig7}
when computing the effective slope. The resulting curve is shown in 
Fig.~\ref{fig9}. 
The transition from  the ``soft'' to the ``hard'' regimes is clearly visible. 
Quite interestingly, $\lambda_{eff}$
at very low $Q^2$ turns out to be close to the standard value of 
$0.08$ for the soft Pomeron. This value is dependent on the choice of 
the quark mass which is not completely constrained. However, if the mass
is chosen such that the cross section in the photoproduction region
roughly agrees with the data it automatically leads to the right value
for the slope in this region.
The other important feature of small-$x$ saturation is that 
with decreasing $\tilde{x}$ the slope $\lambda_{eff}$ also decreases, as 
can be inferred from Fig.~\ref{fig9}. 
In the true high energy asymptotics
it would approach zero which is related to the fact that 
in the complex plane of angular momentum $j$
the leading singularity for our model is located at $j=1$.

\subsection{Charm}\label{sec8}

\begin{figure}
   \vspace*{-1cm}
    \centerline{
     \epsfig{figure=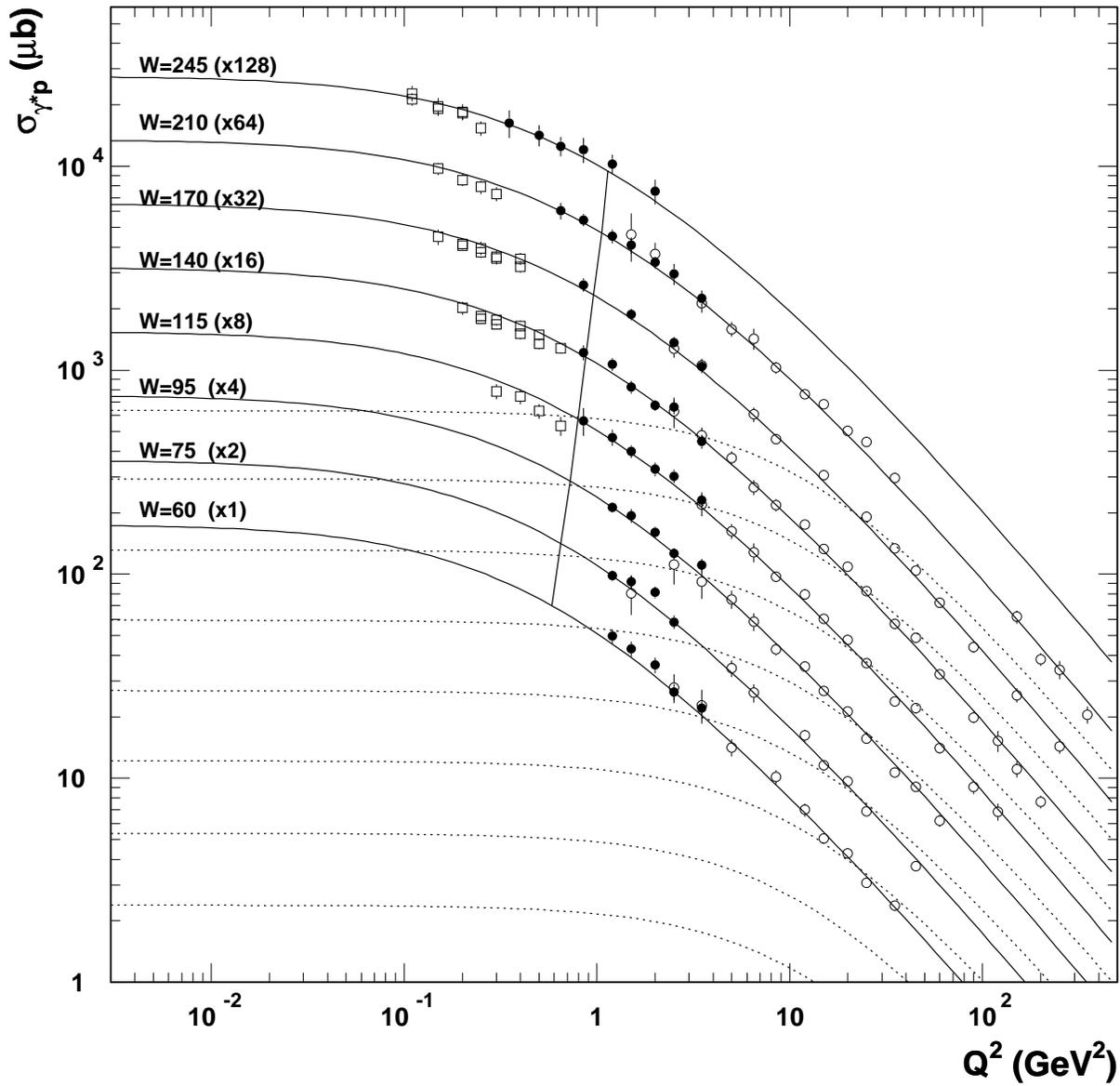,width=18cm}
               }
    \vspace*{-0.5cm}
\caption{The $\gamma^* p$-cross section including charm.
The charm contribution itself is plotted as dotted line.}
\label{fig10}
\end{figure}

Charm gives a substantial contribution to the total cross section
$\sigma^{\gamma^*p}$ and cannot be ignored. 
We have performed a separate fit including
charm but without introducing new parameters. Technically this means that in
the basic formulae (\ref{eq:2}-\ref{eq:4}) the sum over 
flavours has to be extended to include charm with a mass of
$1.5~GeV$. Also effected is  $\tilde{x}$ in the 
corresponding dipole cross section  (\ref{eq:8},\ref{eq:10}) 
since it contains the quark mass (see eq.~(\ref{eq:tech5})).
We  do not change the light flavour mass and keep it equal to  $140~MeV$. 
The impact of charm is shown in the second row of Table.~1 and in 
the last three plots. The $\chi^2$ of the fit has increased
but is still acceptable. This is also reflected in Fig.~\ref{fig10}
in which the agreement with the fitted data looks satisfactory.
The most important change is a shift of the critical line down to smaller
scales (see Fig.~\ref{fig10}).  

\begin{figure}
   \vspace*{-1cm}
    \centerline{
     \epsfig{figure=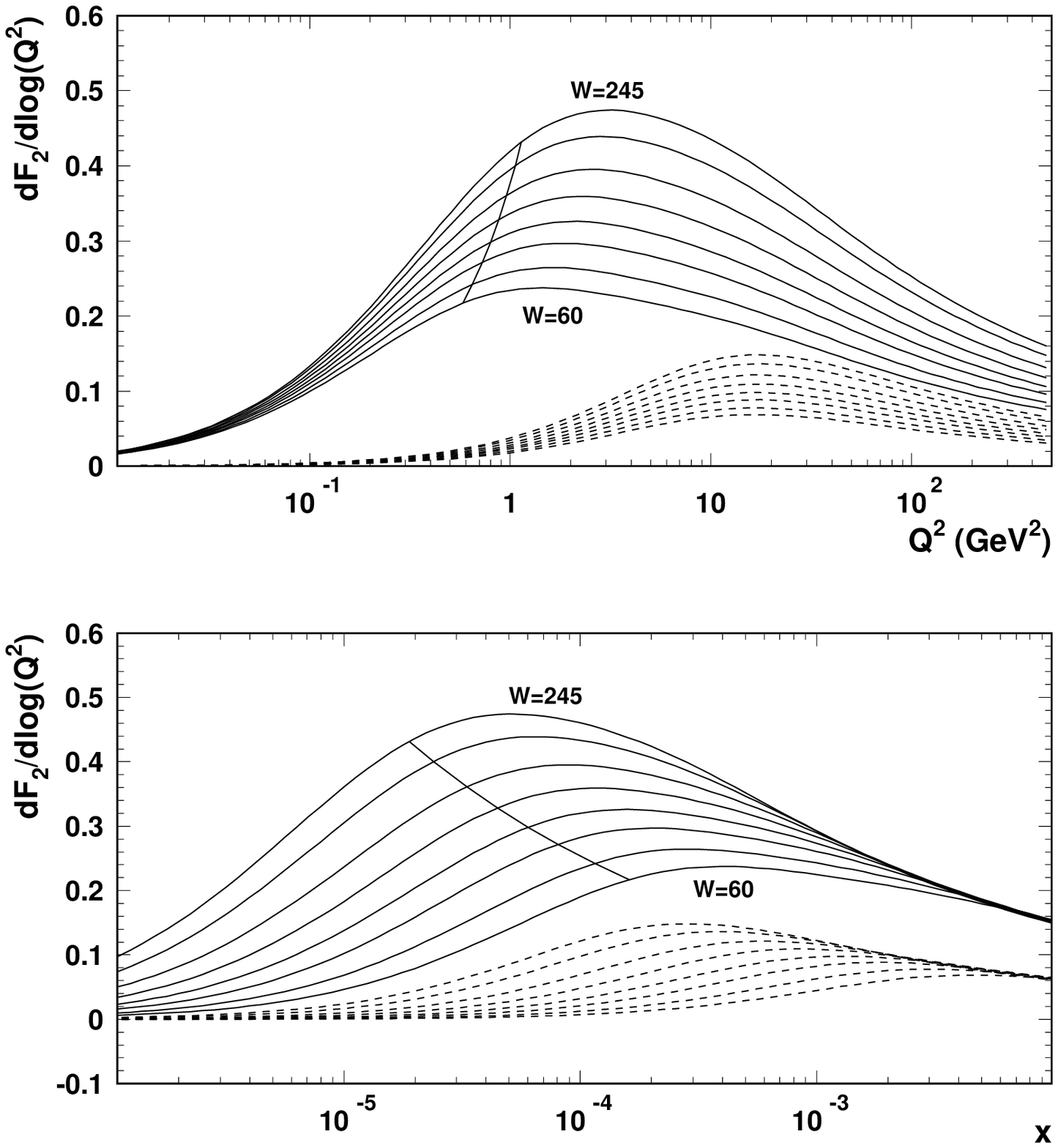,width=18cm}
               }
    \vspace*{-0.5cm}
\caption{The logarithmic $Q^2$-slope of $F_2$ with charm.
The dashed line shows the charm contribution.}
\label{fig11}
\end{figure}

More visible is the effect of introducing charm in the $Q^2$-slope of
$F_2$ shown in  Fig.~\ref{fig11}.
The shape of the peak becomes broader and the value
of the slope at the high end of $Q^2$ and $x$ increases. The main
reason for broadening is the fact that charm saturates at
a much higher scale due to its high mass. We also see that the critical
line moves further away from the maxima. We still believe that the critical
line characterizes the dynamical part of saturation in contrast to the
saturation for charm which is entirely due to its small size and
predominantly perturbative.
But even a small size contribution will at 
extremely large $W^2$  be bigger than the saturation radius $R_0(x)$ 
and undergo small-$x$ saturation.   

A rough comparison with preliminary data on the charmed structure function
shows a reasonable agreement with the data at low $x$. 
We have again looked at the
effective slope $\lambda_{eff}$, shown in Fig.~\ref{fig11},
and find a slight increase
at the low end of $Q^2$ of about $10\%$ as compared to our previous fit
with light flavours only. 

\section{Implication on Diffraction} \label{sec9}

One of the important features of the wave function formalism 
is its flexible applicability for inclusive and
diffractive scattering in DIS. 
We can immediately write down the cross section
for diffractive scattering using the photon wave function defined
in  section \ref{sec2} (see also ref.\cite{NIK2}):
\be\label{eq:diff1}
\left. \frac{d\sigma_{T,L}^D}{d t}\right|_{t=0}\;=\;\frac{1}{16\pi}\;
\int \!d\,^2{\bf r}\! \int_0^1 \!dz \:  
\vert \Psi_{T,L}\,(z,\bfr) \vert ^2 \: |\sigma\,(x,r^2)|^2\;
\\
\ee
and perform the  Mellin transformation as in section \ref{sec3}
\bea
\label{eq:diff2}
\left. \frac{d\sigma_{T,L}^D}{d t}\right|_{t=0}&=&\int \frac{d\nu}{2 \pi}
\int \!d\,^2{\bf r}\! \int_0^1 \!dz \:  
\vert \Psi_{T,L}\,(z,\bfr) \vert ^2 \: \int \frac{dr'^2}{r'^2}
\left(\frac{r}{r'}\right)^{1+2i\nu}|\sigma\,(\tilde{x},r'^2)|^2
\\ \nonumber
\\ \nonumber
&=&\sigma_0^2\;\;
\int \frac{d\nu}{2 \pi} \;\;\left[\left(\frac{x_0}{\tilde{x}}\right)^\lambda
\frac{Q_0^2}{Q^2}\right]^{1/2+i\nu}
\mbox{H$_{T,L}$}\left(\nu,\frac{m_f^2}{Q^2}\right)\;\;\mbox{G}^D(\nu)\;\;.
\eea
The only change appears to be the function G$^D(\nu)$. 
Provided the integrand in the second line of eq.~(\ref{eq:diff2}) has at least
a single pole at $\nu=-i/2$ we find the power behaviour 
with respect to the Bjorken variable $x$ to be the same
for the inclusive as well as diffractive cross section. 
The explicit calculation of the function G$^D$ defined by
\bea \label{eq:diff3}
\mbox{G}^D(\nu)&=&\int_0^\infty d\hat{r}^2\;
\left(\hat{r}^2\right)^{-3/2-i\nu}\;\nonumber
\left(1\,-\,e^{-\hat{r}^2}\right)^2\\ 
&=& \;\left(2^{1/2+i\nu}-2\right)\;\Gamma(-1/2-i\nu)\\ \nonumber
&=& \left(2-2^{1/2+i\nu}\right)\mbox{G}(\nu)
\eea
reveals that G$^D(\nu)$ is almost identical to G$(\nu)$ except for a factor
which gives zero at  $\nu=-i/2$. This zero has a common origin 
for all sensible models and simply reflects the fact that at small distances
$r$ the cross section vanishes proportional to $r^2$ which in diffraction
turns into $r^4$. In section \ref{sec4}, eq.(\ref{eq:tech9}), 
we found a double pole at $\nu=i/2$
in  the integrand for the transverse part 
which is now reduced to a single pole.
Since we are left with at least a single pole, 
the diffractive $\gamma p$-cross section
for transverse polarized photons has indeed the same power behaviour in $x$
as the inclusive $\gamma p$-cross section.

We also found in section \ref{sec4}, eq.(\ref{eq:tech10}), that the inclusive 
longitudinal cross section had only a single pole which
disappears completely in the diffractive cross section.
The conclusion is that the next
pole at $\nu=-3i/2$ becomes the leading pole and the power $1/2+i\nu$ in 
eq.~(\ref{eq:diff2}) equals 2. Consequently, 
the longitudinal part in diffractive scattering rises at small $x$
two times as fast as in the inclusive case. Our  result for the
longitudinal contribution also confirms
the well known higher twist nature of the diffractive cross section.

It is important to note that the results discussed in this section
are derived without imposing cuts on the final state except for
the basic rapidity cut. In particular, there has no cutoff being
assumed for the transverse momentum  of the final state partons. 
This allows   to use formula (\ref{eq:diff1}). 
In ref.~\cite{BLotW} a similar study lead to a closely
related statement about the conservation of ``anomalous dimensions''.
In an extended version with non-zero momentum transfer links can be found
to conformal invariance  in the Regge limit \cite{BLipW}. 
 
We can go a step further and calculate the ratio of the diffractive 
and inclusive cross section in the limit of large $Q^2$. This can be done
by shifting the contour of  the $\nu$-integration
down the lower half of the complex plane,
as was done in section \ref{sec5}, and picking up the leading pole:
\be \label{eq:diff4}
\frac{\sigma^D}{\sigma_{tot}}\;=\;\frac{\sigma_0}{4\pi B}\;
\frac{3\;\ln(2)}{6\;\ln{Q^2}+6\;\lambda\;\ln(x/x_0)+6\;\gamma_E+1}\;,
\ee
where $B$ denotes the diffractive slope.
A mild logarithmic dependence on $x$ is found in the 
ratio which gives a slight logarithmic rise with decreasing $x$. 
Increasing $Q^2$ we find a logarithmic suppression which is due to
the extra zero in  G$^D(\nu)$ at $\nu=-i/2$.
Taking the parameters found in our fit
and the experimental value for $B$ we find that
the ratio turns out to be equal to a few percent. 
We have to note that
a contribution of similar size is expected from the emission of a gluon.
This contribution will have the same power behaviour in $x$ 
as for the quarks and we expect a
ratio of a similar size.  All contributions together will be 
of the order of 10\%.

More precise studies on diffraction based on the model
discussed here  will be presented elsewhere \cite{GW}.

\section{Summary}
\label{sec:8}

We have presented a model which provides a good description
of all DIS data below $x=0.01$ (including the photoproduction 
data). An important ingredient of our approach is the presence
of small-$x$ saturation. This is achieved by introducing an
$x$-dependent saturation radius $R_0(x)=(x/x_0)^{\lambda/2}$ 
which we define with the
help of two parameters. These parameters, together with the overall 
normalization of the cross section, was determined by fits to the 
DIS data. The parameterization obtained in this way was then extended
into the photoproduction region showing reasonable agreement 
with the data there. A particular result of this extrapolation is 
the effective  Pomeron intercept of approximately $1.08$. In the DIS region
this effective intercept increases to the asymptotical value of $1.29$. 
We have introduced a
``critical line'' which marks the boundary of the saturation region
in the $(x,Q^2)$-plane. Finally, we found an intriguing result 
when applying our model to diffractive scattering processes, namely,
an almost constant ratio of diffractive and inclusive DIS cross section.

Our model is basically phenomenological. It is remarkable, however,
that with only three fitted parameters we successfully described the small-$x$
data. The theoretical
basis is the separation of the perturbative
dissociation of the photon into a quark-antiquark pair and 
the subsequent interaction of the pair with the proton.
The latter is modelled in an eikonal way. The main consequence
is that we do not only have saturation at low $Q^2$, 
 but also saturation at low $x$. The natural extension of our model 
would be the incorporation of 
a perturbative treatment for large $Q^2$. Also important
is the development of a microscopical picture behind saturation as proposed
in ref.\cite{Heribert}.

A real test for our model would by a future $ep$-collider. Our prediction is
that with increasing energy the saturation scale moves up in $Q^2$.
For an $1~TeV$ collider it woud be roughly a factor of $2$.

\vskip 1cm
\centerline{\large \bf Acknowledgements}

We thank Alan Martin, Jan Kwiecinski and Misha Ryskin
for valuable discussions.  
KG-B thanks the Royal Society/NATO  for financial support and the Department
of Physics of the University of Durham  for warm hospitality.
This research has been supported in part the Polish State Committee 
for Scientific Research grant no. 2 P03B 089 13 and by the EU
Fourth Framework Programme  
``Training and Mobility of Researchers''  Network,    
``Quantum Chromodynamics and the Deep Structure of Elementary
Particles'', contract   FMRX-CT98-0194 (DG~12-MIHT).


\newpage

\begin{figure}
   \vspace*{-1cm}
    \centerline{
     \epsfig{figure=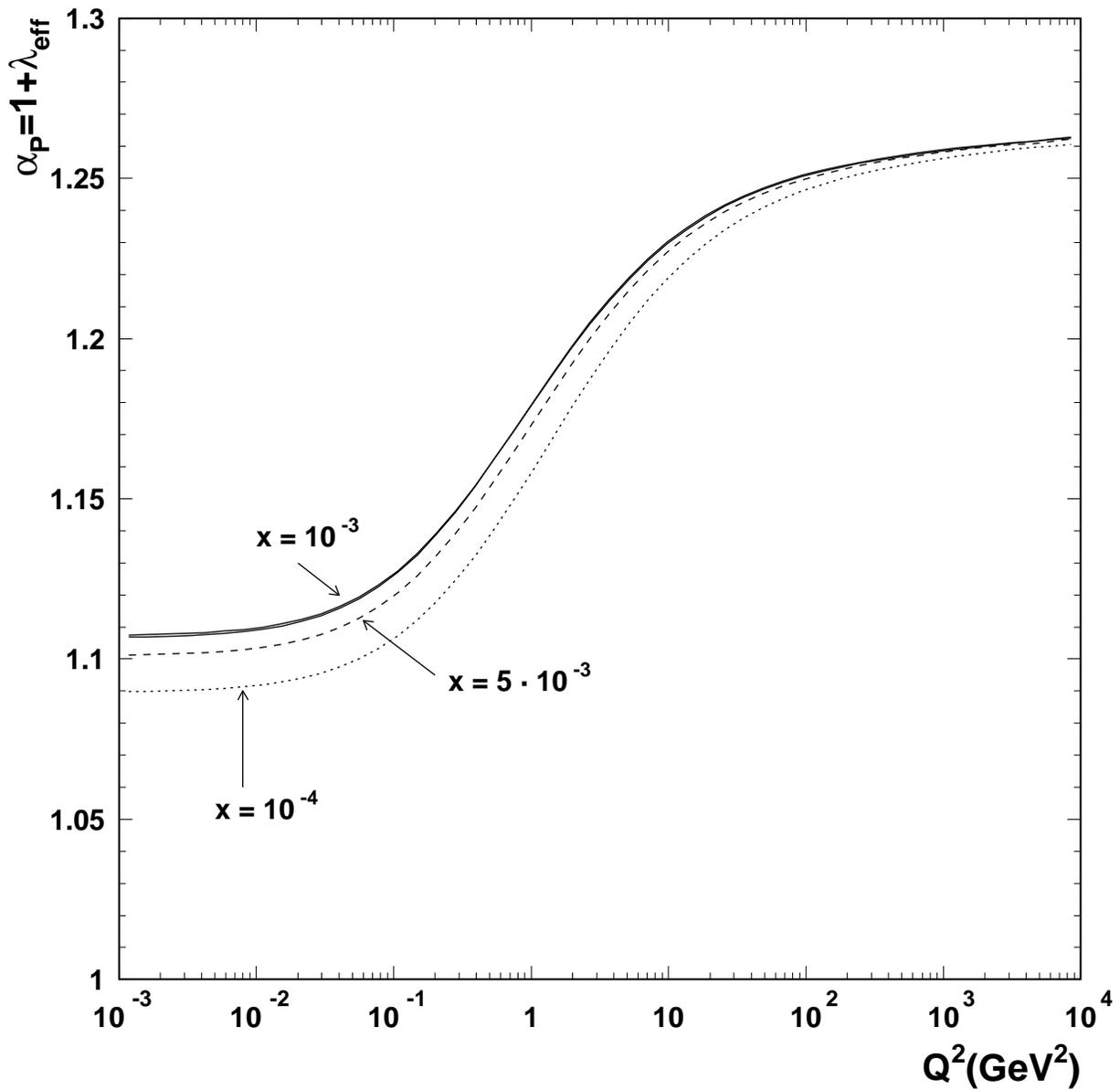,width=18cm}
               }
    \vspace*{-0.5cm}
\caption{The effective slope $\lambda_{eff}$ based on the charm fit. 
The intercept is slightly
higher at low $Q^2$ as compared to fig.\ref{fig9}.}
\label{fig12}
\end{figure} 

\end{document}